\documentclass[10pt, letterpaper]{article}
\usepackage[top=2cm,left=3cm,footskip=0.75in,marginparwidth=2in]{geometry}

\usepackage{natbib}
\usepackage[hidelinks]{hyperref}

\newcommand{\etal}{{\it et al.}}
\newcommand*{\doi}[1]{\href{http://dx.doi.org/#1}{DOI}}

\usepackage{nameref,hyperref}

\makeatletter
\renewcommand{\@biblabel}[1]{\quad#1.}
\makeatother

\usepackage{color}

\definecolor{Gray}{gray}{.25}

\usepackage{graphicx}

\begin{document}
\vspace*{0.35in}

\begin{flushleft}
{\Large
\textbf\newline{Global Gradients for Cosmic-Ray Protons in the Heliosphere During the Solar Minimum of Cycle 23/24}
}
\newline
\\
E.E. Vos\textsuperscript{1,*},
M.S. Potgieter\textsuperscript{1}
\\
\bigskip
$^{\bf{1}}$ Center for Space Research, North-West University, 2520 Potchefstroom, South Africa\\
\bigskip
* etienne.eevos@gmail.com

\end{flushleft}

\section*{Abstract}
Global gradients for cosmic-ray (CR) protons in the heliosphere are computed with a comprehensive modulation model for the recent prolonged solar minimum of Cycle 23/24. Fortunately, the PAMELA (\emph{Payload for Antimatter Matter Exploration and Light-nuclei Astrophysics}) and \emph{Ulysses}/KET (\emph{Kiel Electron Telescope}) instruments simultaneously observed proton intensities for the period between July 2006 and June 2009. This provides a good opportunity to compare the basic features of the model with these observations, including observations from \emph{Voyager}-1 in the outer heliosphere, beyond $50\,$AU. 
Radial and latitudinal gradients are calculated from measurements, with the latter possible because \emph{Ulysses} changed its position significantly in the heliocentric meridional plane during this period.
The modulation model is set up for the conditions that prevailed during this unusual solar minimum period to gain insight into the role that particle drifts played in establishing the observed gradients for this period.
Four year-end PAMELA proton spectra were reproduced with the model, from 2006 to 2009, followed by corresponding radial profiles that were computed along the \emph{Voyager}-1 trajectory, and compared to available observations. 
It is found that the computed intensity levels are in agreement with solar minimum observations from \emph{Voyager}-1 at multiple energies. The model also reproduces the steep intensity increase observed when \emph{Voyager}-1 crossed the heliopause region. 
Good agreement is found between computed and observed latitudinal gradients so that we conclude that the model gives a most reasonable representation of modulation conditions from the Earth to the heliopause for the period from 2006 to 2009.
As a characteristic feature of CR drifts, the most negative latitudinal gradient is computed for 2009, with a value of $-0.15\,$\%\,degree$^{-1}$ around $600\,$MV. 
The maximum radial gradient in the inner heliosphere (as covered by \emph{Ulysses}) also occurs in this range, with the highest value of $4.25\,$\%\,AU$^{-1}$ in 2009.
\\
\vskip 0.04cm
\noindent\textbf{Keywords}:  {Cosmic Rays, Galactic;  Solar Cycle, Models;  Energetic Particles, Propagation.}

\section{Introduction}
The \emph{Ulysses}/KET (\emph{Kiel Electron Telescope}) and PAMELA (\emph{Payload for Antimatter Matter Exploration and Light-nuclei Astrophysics}) missions overlapped between July 2006 and June 2009, measuring galactic proton intensities at Earth and along \emph{Ulysses}' orbit during most of the unusual solar minimum of Cycle 23/24.  Together with \emph{Voyager}-1 observations in the outer heliosphere during this period, this created an opportunity to study the global radial and latitudinal gradients with a comprehensive three-dimensional (3D) modulation model. Examining the main features of this model and relating them to the mentioned observations is the main purpose of this study. The model was applied previously to reproduce proton and electron spectra from PAMELA from mid-2006 to the end of 2009 so that the modulation conditions and parameters are set for this period as described in detail by \citeauthor{PotgieterVosetal2014} (\citeyear{PotgieterVosetal2014}, \citeyear{PotgieterVosetal2015}) and \citet{VosPotgieter2015}. For an independent modelling investigation of this unusual period, see \citet{Zhaoetal2014}.
\par
Numerical models can be used to compute intensity gradients at any position in the heliosphere. These gradients are known as local gradients and can unfortunately not be used to compare with gradients calculated from observations, which are mostly far apart in terms of radial and meridional distances (\citeauthor{Potgieteretal1989}, \citeyear{Potgieteretal1989}).  Such gradients are usually a mixture of radial and latitudinal gradients and are known as global gradients.  These are calculated from differential intensities between, for example, the position of Earth and that of \emph{Ulysses} in the inner heliosphere, or \emph{Voyager}-1 beyond $50\,$AU \citep{WebberandLockwood1986, Cummingsetal1987, Mcdonaldetal1997, Heberetal1998}.
\par
Numerical models are easily adjusted to compute global gradients, exactly as is done with observational data (\emph{e.g.} \citeauthor{NgobeniPotgieter2011}, \citeyear{NgobeniPotgieter2011}). To do so, the basic approach of \citet{Desimoneetal2011}, \citet{Gieseleretal2013}, and \citet{GieselerHeber2016} is followed, which makes for an appropriate comparison between the computed and observational gradients.  This study also provides insight into the role that particle drifts had played during the unusual minimum period of Cycle 23/24, in particular why the recent gradients seem smaller than what was predicted for this magnetic polarity cycle; see the reviews by \citet{McKibben2005}, \citet{HeberPotgieter2006}, and \citet{Potgieter2014b}.
\par
First, the numerical model is briefly discussed together with the essential modulation parameters. This is followed by a discussion of the reproduced PAMELA proton spectra \citep{Adrianietal2013} and the computed radial-intensity profiles from the Earth to the heliopause (HP) for the period from 2006 to 2009.  This is compared to \emph{Voyager}-1 observations as it moved through the outer heliosphere, for kinetic energies [$E$] between $142\,$MeV and $215\,$MeV \citep{Webberetal2011}.  In the following sections, we discuss how the global spatial gradients are computed based on cosmic-ray (CR) intensities from PAMELA at the Earth and along the orbit of \emph{Ulysses}. The computed radial and latitudinal gradients are presented as a function of rigidity, and it is shown how these spatial gradients developed from 2006 to 2009. The computed gradients from the model are then compared to corresponding calculated observational values.

\section{Modelling Parameters}
A comprehensive 3D modulation model is used to compute differential intensities of protons at the Earth and throughout the heliosphere, and it is based on the numerical solution of the transport equation from \citet{Parker1965}:
\newcommand{\qvec}[1]{\textbf{\textit{#1}}}
\begin{eqnarray}
  \frac{\partial f}{\partial t} &=& -\left(\qvec{V} + \left<\qvec{v}_D\right>\right)\cdot\nabla f + \nabla\cdot\left(\qvec{K}_s\cdot\nabla f\right)\nonumber\\
  && + \frac{1}{3}\left(\nabla\cdot\qvec{V}\right)\frac{\partial f}{\partial \ln P},
\end{eqnarray}
with $f$ the particle distribution function, and $t$ the time where $\partial f/\partial t=0$, since we address modulation during solar minimum when modulation parameters change gradually.  The terms on the right-hand side represent convection, with $\qvec{V}$ the solar wind (SW) velocity;  averaged particle drift velocity $\left<\qvec{v}_D\right>$ caused by gradients, curvatures, and heliospheric current sheet (HCS) drifts in the global heliospheric magnetic-field (HMF);  diffusion, with $\qvec{K}_S$ the symmetric diffusion tensor; then adiabatic energy losses, with $P$ the rigidity in GV.
\par
Contained within $\qvec{K}_S$ are three diffusion coefficients (DCs) as discussed below. With the DCs related to the mean free paths [MFPs; $\lambda$] by $\kappa=\lambda\left(v/3\right)$, where $v$ is the particle speed, the equation for the DC parallel to the average background HMF, at a radial distance of $r$, polar angle $\theta$ and azimuthal angle $\phi$, is given in general by
\begin{equation}\label{eq:kappa_parallel}
  \kappa_\parallel = \kappa_{\parallel 0}\,\beta\,F\left(r,\theta, \phi\right)\,G\left(P\right),
\end{equation}
with
\begin{equation}
  F\left(r,\theta, \phi\right) = \frac{B_0}{B},
\end{equation}
and
\begin{equation}\label{eq:G(P)}
  G\left(P\right) = \left(\frac{P}{P_0}\right)^a\left\{\frac{\left(\frac{P}{P_0}\right)^c+\left(\frac{P_k}{P_0}\right)^c}{1+\left(\frac{P_k}{P_0}\right)^c}\right\}^{\frac{b-a}{c}}.
\end{equation}
Here $\kappa_{\parallel 0}$ is a scaling constant in units of cm$^2$\,s$^{-1}$, $B$ the magnetic-field magnitude in nT, $\beta$ the ratio of particle speed to the speed of light, and $P$ as before.  The variables $a$, $b$, $c$, and $P_k$ determine the shape of the rigidity dependence of the MFP, which has the functional form of two power laws combined; see \citet{VosPotgieter2015} for a list of their values.  The constants $B_0=1\,$nT and $P_0=1\,$GV keep $F$ and $G$ dimensionless.
\par
The expression for the HMF is modified according to \citet{SmithBieber1991} and is given by
\begin{equation}
  B = B_n\,\left(\frac{r_0}{r}\right)^2\,\sqrt{1 + \tan{\psi}^2},
\end{equation}
with $r_0=1\,$AU, and $\tan{\psi}$ a function of $r$ and $\theta$ given by
\begin{equation}
  \tan{\psi} = \frac{\Omega\,\left(r - r_{sb}\right)\,\sin{\theta}}{V(r,\theta)} - \frac{r\,V\left(r_{sb}, \theta\right)}{r_{sb}\,V\left(r,\theta\right)}\,\,\frac{B_T\left(r_{sb}\right)}{B_R\left(r_{sb}\right)},
\end{equation}
where $B_n$ is a normalization constant that ensures that the HMF has the value $B_e$ at Earth, $\Omega$ the angular velocity of the Sun, and $V$ the SW speed.  With the solar radius $R_\odot=0.005\,$AU, the value $r_{sb}=10\,R_{\odot}$ and the ratio $B_T/B_R=0.15$ are constants that determine the HMF modification. For a complete description and study of these types of modifications, see \citet{Raathetal2016}.
\par
For diffusion perpendicular to the HMF, a rigidity dependence similar to that of $\lambda_\parallel$ is assumed below {4\,}{GV}, but with a slightly weaker dependence of $P^{1.58}$ above {4\,}{GV}.  This is in line with what is required from basic turbulence theory (see also \citeauthor{Straussetal2013}, \citeyear{Straussetal2013}, and \citeauthor{Manueletal2014}, \citeyear{Manueletal2014}).  Following \citet{VosPotgieter2015}, a distinction is also made between the radial [$\kappa_{\perp r}$] and polar [$\kappa_{\perp \theta}$] diffusion by scaling the former by $2\,$\% of parallel diffusion and the latter by $1\,$\%.    
These DCs are given by
\begin{equation}\label{eq:kappa_perp_r}
  \kappa_{\perp r} = 0.02\,\kappa_{\parallel 0}\,\beta\,F\left(r,\theta, \phi\right)\,G_{\perp}\left(P\right)
\end{equation}
and
\begin{equation}\label{eq:kappa_perp_theta}
  \kappa_{\perp \theta} = 0.01\,\kappa_{\parallel 0}\,\beta\,F\left(r,\theta, \phi\right)\,G_{\perp}\left(P\right)\,H_{\perp \theta},
\end{equation}
with $G_{\perp}$ also generally defined by Equation (\ref{eq:G(P)}). Here,
\begin{equation}\label{eq:h_perp_theta}
  H_{\perp \theta} = A^+ \mp A^-\,\tanh{\left[8\left(\theta_A-90^\circ\pm\theta_F\right)\right]},
\end{equation}
with $A^\pm = (3\pm1)/2$, $\theta_F=35^\circ$, $\theta_A=\theta$ for $\theta\le 90^\circ$ but $\theta_A = 180^\circ-\theta$ for $\theta > 90^\circ$.  
Equation (\ref{eq:h_perp_theta}) enhances $\kappa_{\perp \theta}$ towards the heliospheric polar regions as motivated by \citet{Potgieter2000}; see also related discussions by \citeauthor{NgobeniPotgieter2011} (\citeyear{NgobeniPotgieter2011}, \citeyear{NgobeniPotgieter2014}). 
\par
\citet{PotgieterVosetal2014} illustrated that particle drifts contributed to the high intensities observed by PAMELA in 2009. The rigidity and spatial dependence for the drift coefficient that they used is given by
\begin{equation}\label{eq:kappa_a}
  \kappa_A = \frac{\beta P}{3B}\,\frac{\left(\frac{P}{P_{A 0}}\right)^2}{1 + \left(\frac{P}{P_{A 0}}\right)^2},
\end{equation}
which reduces drifts below $P_{A 0} = 0.55$\,GV with respect to the weak scattering case, which is simply proportional to $\beta P$ \citep[for an elaborate discussion on this topic, see][]{NgobeniPotgieter2015}.  
This is required to explain the small latitudinal gradients observed by \emph{Ulysses} at low rigidities; see also \citet{Heberetal1997}, \citet{HeberPotgieter2006}, and for supportive theoretical discussions, see \citet{Burgeretal2000}.
\par
Figure 1 shows the rigidity dependence of the four sets of MFPs and the drift scales [AU] as applied in the model to reproduce the aforementioned PAMELA spectra.  
\begin{figure}[t]
  \begin{center}
  \includegraphics[width=0.75\linewidth]{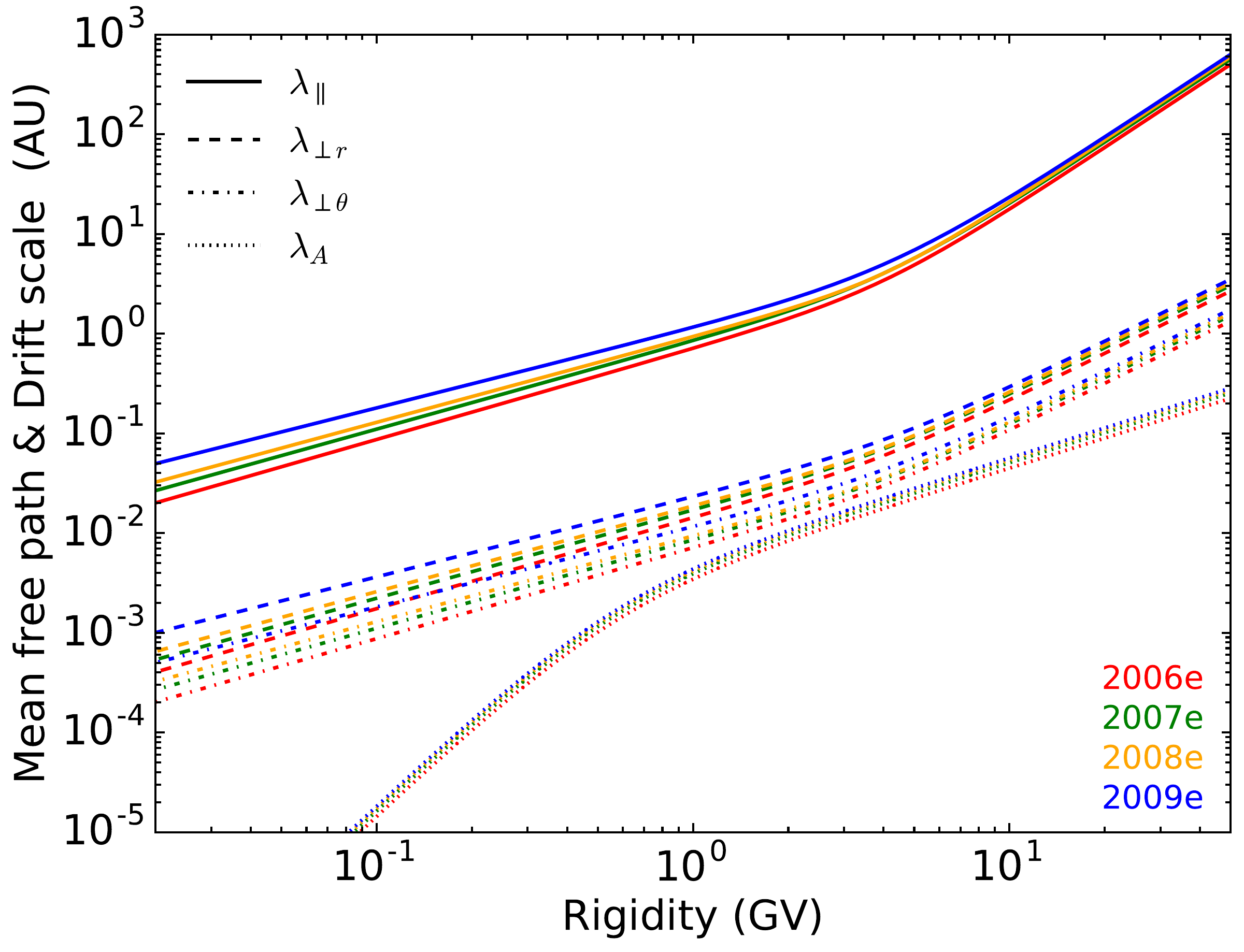}
  \end{center}
  \caption{The rigidity dependence of the proton MFPs for diffusion parallel and perpendicular to the magnetic-field lines at the Earth, and for the drift scale [AU].  Parallel MFPs [$\lambda_{\parallel}$] are given by the solid lines, while perpendicular MFPs in the radial [$\lambda_{\perp r}$] and polar [$\lambda_{\perp\theta}$] directions are given by the dashed and dashed--dotted lines, respectively.  The drift scale [$\lambda_A$] is given by the dotted lines.  See Equations (\ref{eq:kappa_parallel}), (\ref{eq:kappa_perp_r}), (\ref{eq:kappa_perp_theta}), and (\ref{eq:kappa_a}).
	\label{fig:prot_dc}}
\end{figure}
Furthermore, the HCS tilt angle [$\alpha$] and the HMF at Earth [$B_e$] are modulation parameters that influence both diffusion and drift.  As an attempt to account for the time-dependent changes that occur in these parameters while using a steady-state model, realistic modulation conditions for each of the selected 27-day averaged PAMELA spectra were set up in the model.  These aspects are described and explained in detail by \citet{VosPotgieter2015}.
\par
Additionally, because of the dynamic nature of the global heliosphere \citep[\emph{e.g.}][]{RichardsonWang2011}, the position of the TS is also varied with solar activity in the model. Since this expands or shrinks the heliosheath, it affects the modulated intensities in the outer heliosphere, and even slightly at Earth; see \citet{Manueletal2015} for an example of the effects of an oscillating heliosheath width on CRs. The HP is positioned at 122 AU in the model to make it consistent with the reported \emph{Voyager}-1 observation (\citeauthor{Stoneetal2013}, \citeyear{Stoneetal2013};  \citeauthor{WebberMcdonald2013}, \citeyear{WebberMcdonald2013}). In our model the DCs are reduced inside the heliosheath, but with the same spatial dependence as in the rest of the heliosphere, except when the HP is approached, where they decrease exponentially to account for the upward jumps in the CR intensity across the HP region observed by \emph{Voyager}-1 \citep[][]{WebberQuenby2015}. This is illustrated in the next section.
Adiabatic energy losses in the heliosheath are treated as in the rest of the heliosphere. By accounting for all of the above variables in the model, and carefully adjusting the DCs, each consecutive year-end PAMELA proton spectrum was reproduced satisfactorily, as we show next. Reproducing the mentioned PAMELA spectra is the departure point of the rest of our investigation.

\begin{figure}[t]
  \begin{center}
    \includegraphics[width=0.75\linewidth]{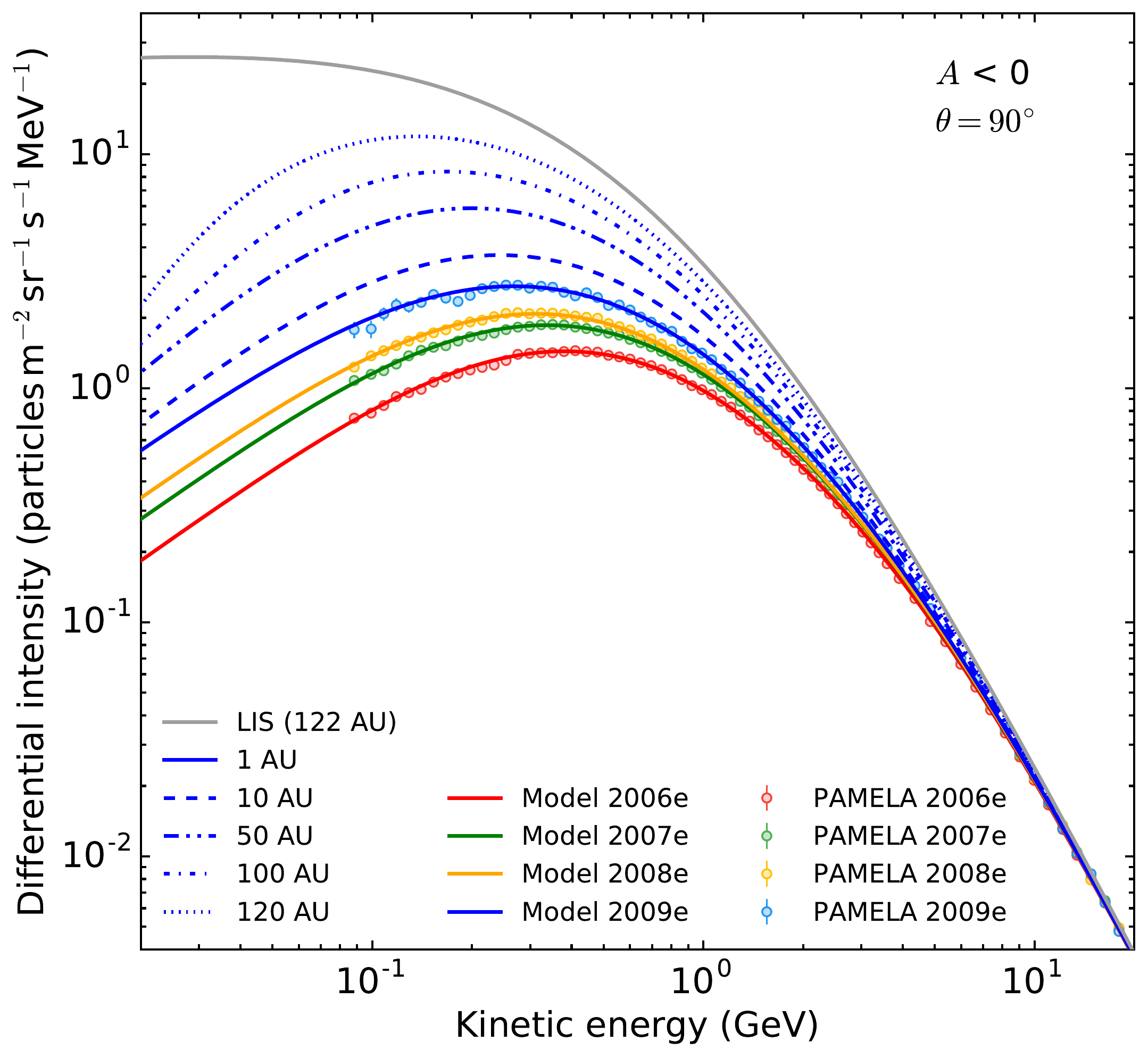}
  \end{center}
  \caption{The observed {PAMELA} and computed proton spectra at the Earth [1 AU] for the periods 2006e, 2007e, 2008e, and 2009e as described in the text.  Corresponding computed spectra are also shown in the equatorial plane (polar angle $\theta=90^\circ$) at {10\,}{AU}, {50\,}{AU}, {100\,}{AU}, and {120\,}{AU} by the blue lines. The LIS from \citet{VosPotgieter2015}, given by the grey line, is specified in the model at {122\,}{AU}.\label{fig:prot_es_fit}}
\end{figure}

\section{The PAMELA Proton Spectra}
Four 27-day averaged PAMELA proton spectra are reproduced, namely for November 2006, December 2007, December 2008, and December 2009, referred to as the 2006e, 2007e, 2008e, and 2009e spectra, respectively, where the suffix ``e'' denotes the end of the year. These are shown in \mbox{Figure 2}, along with the computed spectra. The model reproduces the behaviour of  these galactic proton spectra at all energies of interest to solar modulation studies. It is also clear that the proton spectrum became progressively softer from 2006 to 2009, reaching a maximum intensity of $2.7\,${particles$\,$m$^{-2}\,$sr$^{-1}\,$s$^{-1}\,$MeV$^{-1}$} at the end of 2009, with an accompanying shift of the spectrum peak down to $270\,$MeV.
Below $100\,$MeV the modulated spectra bend into the characteristic $E^1$ slope as a result of adiabatic energy losses caused by the expanding SW, which is the dominant process at non-relativistic proton energies (below $100\,$MeV -- $200\,$MeV).  Predictions of intensity levels can be made with the model for energies below {80\,}{MeV}, where PAMELA measurements are unavailable. At $10\,$MeV, the 2006e intensity is estimated at $\approx\,0.09\,$ particles$\,$m$^{-2}\,$sr$^{-1}\,$s$^{-1}\,$MeV$^{-1}$ increasing to $\approx\,0.3\,$ particles$\,$m$^{-2}\,$sr$^{-1}\,$s$^{-1}\,$MeV$^{-1}$ at the end of 2009. 
\par
Corresponding computed spectra in the equatorial plane of the heliosphere (polar angle/co-latitude of $\theta=90^\circ$) are also shown at {10\,}AU, {50\,}AU, {100\,}AU, and {120\,}AU in relation to the newly constructed very local interstellar spectrum (vLIS) for protons from \citet{VosPotgieter2015}, specified at {122\,}AU where the HP is located.  See also \citet{Potgieter2014a} for new LISs that are based on PAMELA and \emph{Voyager}-1 observations. The computed spectra at larger radial distances show how the maximum in the spectra progressively shifts to lower energies. Protons below {100\,}MeV  experience a significant amount of modulation, even at large radial distances close the HP.
The displayed radial dependence of the spectra is discussed in more detail below. 

\section{Computed Radial Profiles for 2006 to 2009}
The focus in this section is on the spatial distribution of proton intensities, which is investigated to obtain a better global view of the behaviour of CRs throughout the heliosphere during the unusual minimum-modulation conditions of Cycle 23/24.
\par
\begin{figure}[p!]
  \begin{center}
    \includegraphics[width=0.9\linewidth]{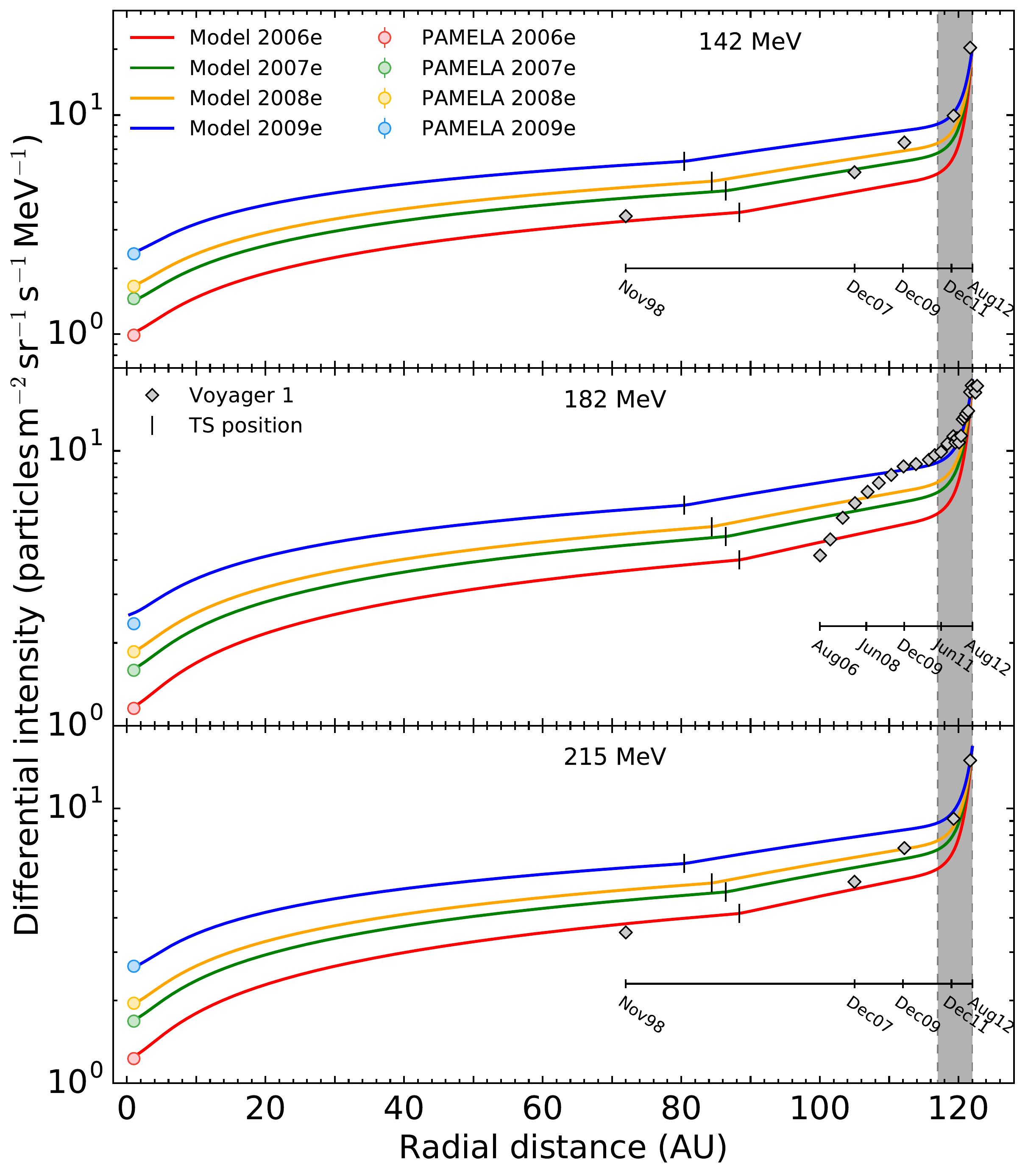}
  \end{center}
  \caption{Modelled radial proton intensities at $56^\circ$ latitude between 2006e (red line) and 2009e (blue line) at energies between $142\,$MeV and $215\,$MeV (from top to bottom panel) are compared to \emph{Voyager}-1 measurements (diamonds; from \citeauthor{Webberetal2011}, \citeyear{Webberetal2011}; \citeauthor{Webberetal2013}, \citeyear{Webberetal2013}).  The panel legend applies to all panels. The TS positions are indicated by the short vertical black lines, with the PAMELA measurements given by the coloured circles at $1\,$AU. A timescale for the \emph{Voyager}-1 data points is inserted in each panel. The significance of the shaded region beyond $116\,$AU is discussed in the text.\label{fig:prot_radial_int}}
\end{figure}
In \mbox{Figure 3} four snapshots of the computed radial-intensity profiles are shown at $56^\circ$ latitude (polar angle of $34^\circ$) to coincide with the \emph{Voyager}-1 trajectory. The radial profiles are shown at energies of {142\,}{MeV}, {182\,}{MeV}, and {215\,}{MeV} (from top to bottom) to coincide with \emph{Voyager}-1 measurements. These profiles are based on the mentioned computed (modelling) spectra for the periods 2006e, 2007e, 2008e, and 2009e, shown here as coloured circles at {1\,}{AU}.  It is assumed that the position of the TS shifted closer to the Sun over these four years as indicated by short vertical black lines, in accord with \citet{Manueletal2014}, who based their modelling on the observations reported by \citet{Stoneetal2013}.

\emph{Voyager}-1 proton observations from \citet{Webberetal2013} are shown for different periods by the grey diamonds. 
Evidently, the radial dependence of the intensity profiles changes at and beyond the TS and is even more significant close to the HP. These aspects are discussed below in more detail.

\begin{figure}[t!]
  \begin{center}
    \includegraphics[width=0.85\linewidth]{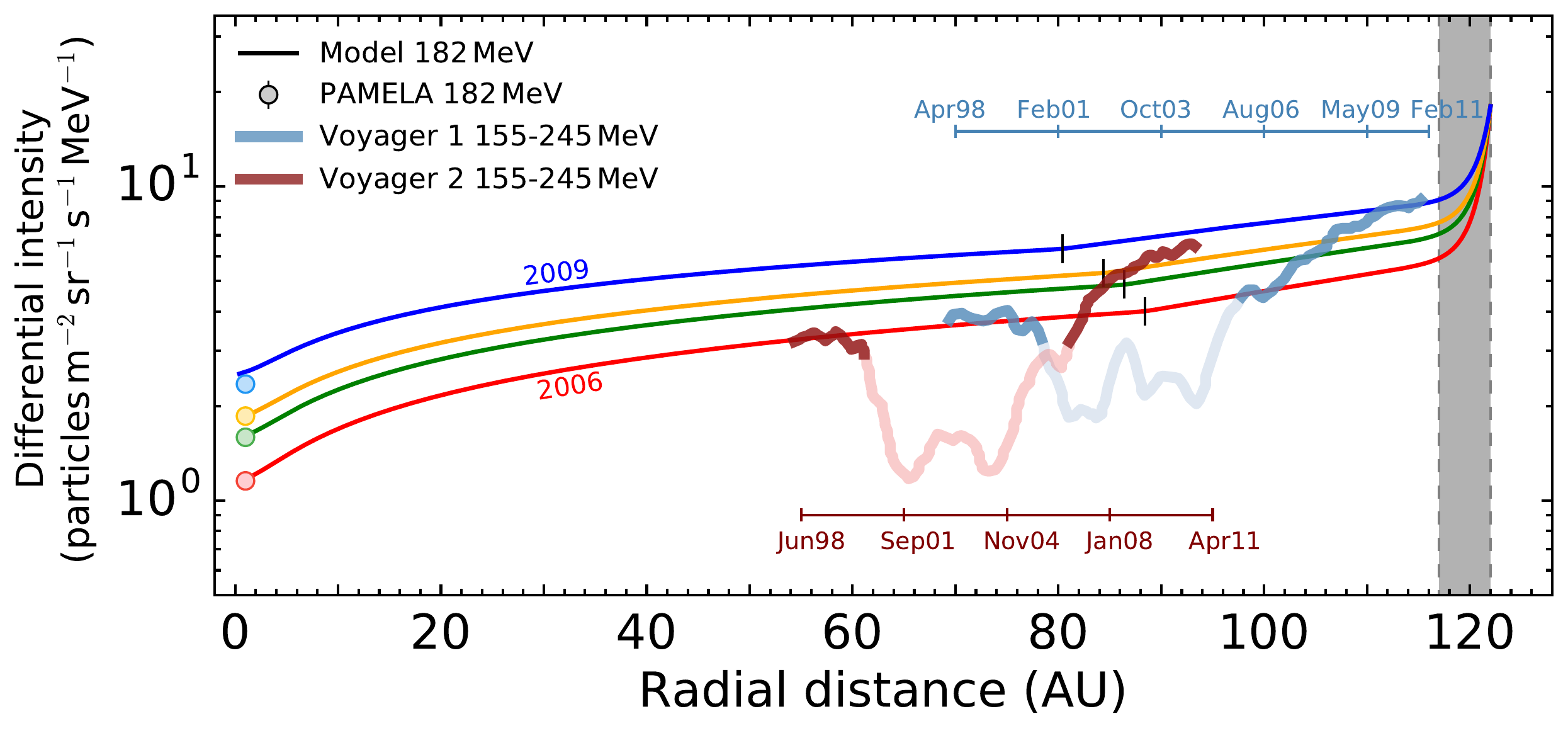}
  \end{center}
  \caption{Modelled radial intensities at $56^\circ$ latitude between 2006e (red line) and 2009e (blue line) for $182\,$MeV protons are compared to \emph{Voyager}-1 and -2 measurements (given by the thick blue and maroon lines, respectively) in the $155\,$MeV -- $245\,$MeV energy range \citep{Webberetal2011}.  Dark line segments represent \emph{Voyager} intensities measured during solar minimum, while faint line segments represent intensities from solar maximum.  The TS positions are indicated by the short vertical black lines, with the PAMELA measurements given by the coloured circles at $1\,$AU.  A time-line is given for each set of \emph{Voyager} measurements.  The significance of the shaded region beyond $116\,$AU is discussed in the text.\label{fig:prot_radial_int2}}
\end{figure}
During the peak of the 1997 solar minimum, \emph{Voyager}-1 was at {72\,}{AU} and observed intensities of $\approx\,${3.5\,}{particles$\,$m$^{-2}\,$sr$^{-1}\,$s$^{-1}\,$MeV$^{-1}$} at {142\,}{MeV}.  According to the model, such intensities are expected to have existed already in the beginning of the recent solar minimum, as shown by the red lines in \mbox{Figure 3} for 2006e.  Observations during December 2007, at {105\,}{AU}, and December 2009, at {112\,}{AU}, in the top and bottom panels are slightly lower than the computed intensities.  This is expected since \emph{Voyager}-1 at this distance essentially measured intensities that correspond to earlier modulation conditions closer to the Sun.  Measurements of {182\,}{MeV} protons beyond {100\,}{AU} (middle panel) are consistent with the modelling results, showing the increase in intensities with increasing distance.
\par
Over the last $\approx\,${5\,}{AU} region in front of the HP (indicated by the shaded region), \emph{Voyager}-1 measured a pronounced increase in intensities \citep[\emph{e.g.}][]{WebberQuenby2015} when it entered the so-called heliosheath depletion region, now recognized as the heliopause region, after which it started to observe the vLIS from August 2012 onward (\citeauthor{Gurnettetal2013}, \citeyear{Gurnettetal2013}).  This jump-like increase is interpreted as a particle barrier-region over which the SW begins to interact with the very local interstellar medium (LISM) causing CRs to jump up to the very local interstellar values over a relatively short distance. This jump-effect is also simulated and discussed by \citet{Luoetal2015}, who considered how the MFPs should scale over this region; obviously, the smaller DCs inside the heliosheath must increase sharply across the HP to match the very LISM values as shown by them.
\par
In \mbox{Figure \ref{fig:prot_radial_int2}} additional \emph{Voyager}-1 and -2 measurements for {155\,}{MeV} -- {245\,}{MeV} protons are shown as a function of radial distance from 1998 to 2011, including the 2009 minimum \citep{Webberetal2011}.  Since observations between 1997 and 2000 were made during an $A > 0$ polarity cycle, they are expected to be lower than intensities from an $A < 0$ cycle, which is the case when comparing the computed radial profiles with \emph{Voyager}-1 measurements at radial distances of {53\,}{AU} -- {61\,}{AU} and \emph{Voyager}-2 observations at {68\,}{AU} -- {78\,}{AU}. The difference in the computed proton fluxes between these two trajectories is quite small, so that the \emph{Voyager}-2 profiles are not plotted. 
The development of proton intensities during the maximum of Solar Cycle 23 can be followed by the fainter segments of these lines.  
\par
Solar minimum observations from the $A < 0$ polarity cycle between 2006 and 2011 are shown at distances of {97\,}{AU} -- {115\,}{AU} and {81\,}{AU} -- {93\,}{AU} for \emph{Voyager}-1 and -2, respectively.  Evidently, the observed radial profiles, set into the right context, are consistent with the computed profiles for the period 2006 to 2009. It seems reasonable to claim that these computed radial profiles for this period are a good representation of the galactic CR intensities throughout the heliosphere during the unusual solar minimum period of Cycle 23/24.

\section{Spatial Gradients in the Inner Heliosphere}
Gradient and curvature drifts of CRs in the heliosphere are recognized as the pre-eminent driving force behind the 22-year CR solar cycle.  The effects of drifts on galactic CRs during different polarity epochs can be measured in the radial [$G_r$] and latitudinal [$G_\theta$] global gradients, which give distinct, but not invariant, results for each solar cycle; characteristically, drifts cause negative latitudinal gradients during $A < 0$ epochs \citep{Potgieteretal2001}.
\par
\citet{Desimoneetal2011} used measurements from PAMELA and \emph{Ulysses}/KET to investigate the radial and latitudinal gradients of protons in the inner heliosphere during the $A < 0$ solar minimum leading up to 2009.  For the rigidity interval $1.6\,$GV -- $1.7\,$GV, they found a radial gradient of $(2.7 \pm 0.2)\,$\%\,AU$^{-1}$ and a latitudinal gradient of $(-0.024 \pm 0.005)\,$\%\,degree$^{-1}$, with the latter less negative than the large negative latitudinal gradients predicted by earlier drift models \citep{Heberetal1997, Potgieteretal2001}. In hindsight, this was an indication that the modulation conditions during the minimum of Cycle 23/24 were different and not indicative of drifts being overall less important in heliospheric modulation, as discussed by \citet{PotgieterVosetal2014}.
\par
This study aims to calculate $G_r$ and $G_\theta$ for the recent solar minimum using the computed intensities as described above. In addition, for comparison with PAMELA and \emph{Ulysses}/KET observations, a pragmatic, empirical method is used to calculate these  observational global gradients, similar to \citet{Desimoneetal2011} and \citet{Gieseleretal2013}; see also the very recent work of \citet{GieselerHeber2016}. As mentioned above, numerical models can be used to compute very precise local radial and latitudinal gradients for a given position anywhere in the heliosphere, which illustrate the unique modulation characteristics of drifts (\citeauthor{Potgieteretal1989}, \citeyear{Potgieteretal1989}; \citeauthor{NgobeniPotgieter2011}, \citeyear{NgobeniPotgieter2011}). However, these local gradients are not useful if a meaningful comparison needs to be made with observations at vastly different locations in the heliosphere.

\subsection{An Empirical Approach}
It is assumed that the temporal and spatial variations of CR intensities are separable in time and space \citep{Mcdonaldetal1997}.  Let $J_U(P,t,r_U,\theta_U)$ be the CR intensity at rigidity $P$ and time $t$ along \emph{Ulysses}' orbit, with a heliocentric distance of $r_U$ and a latitude of $\theta_U$, averaged over one solar rotation.  The intensity at PAMELA  [$J_E(P,t,r_E,\theta_E)$], with $r_E$ and $\theta_E$ the radial distance and latitude of Earth respectively, can be related to $J_U$ with a function $g(P,\Delta{r},\Delta{\theta})$, so that
\begin{equation}\label{eq:JuJeG}
  J_U(P,t,r_U,\theta_U) = J_E(P,t,r_E,\theta_E)\, g(P,\Delta{r},\Delta{\theta}),
\end{equation}
where $g$ depends on the differences in heliocentric distance [$\Delta{r}=r_U-r_E$] and latitude [$\Delta\theta=\left|\theta_U\right|-\left|\theta_E\right|$] between \emph{Ulysses} and PAMELA at Earth.  For simplicity, the Earth's orbital inclination with respect to the solar Equator is ignored so that $r_E=1.0\,$AU and $\theta_E=0^\circ$.  \mbox{Figure 5} gives the heliocentric distance and latitude of \emph{Ulysses} between 2006 and mid-2009 (top panel), as well as the quantities $\Delta{r}$ and $\Delta\theta$ (bottom panel).  A symmetric distribution of CRs is assumed with respect to the heliographic Equator, neglecting possible small asymmetries found by \emph{e.g.} \citet{Heberetal1996b}.

\begin{figure}
  \begin{center}
  \includegraphics[width=0.8\linewidth]{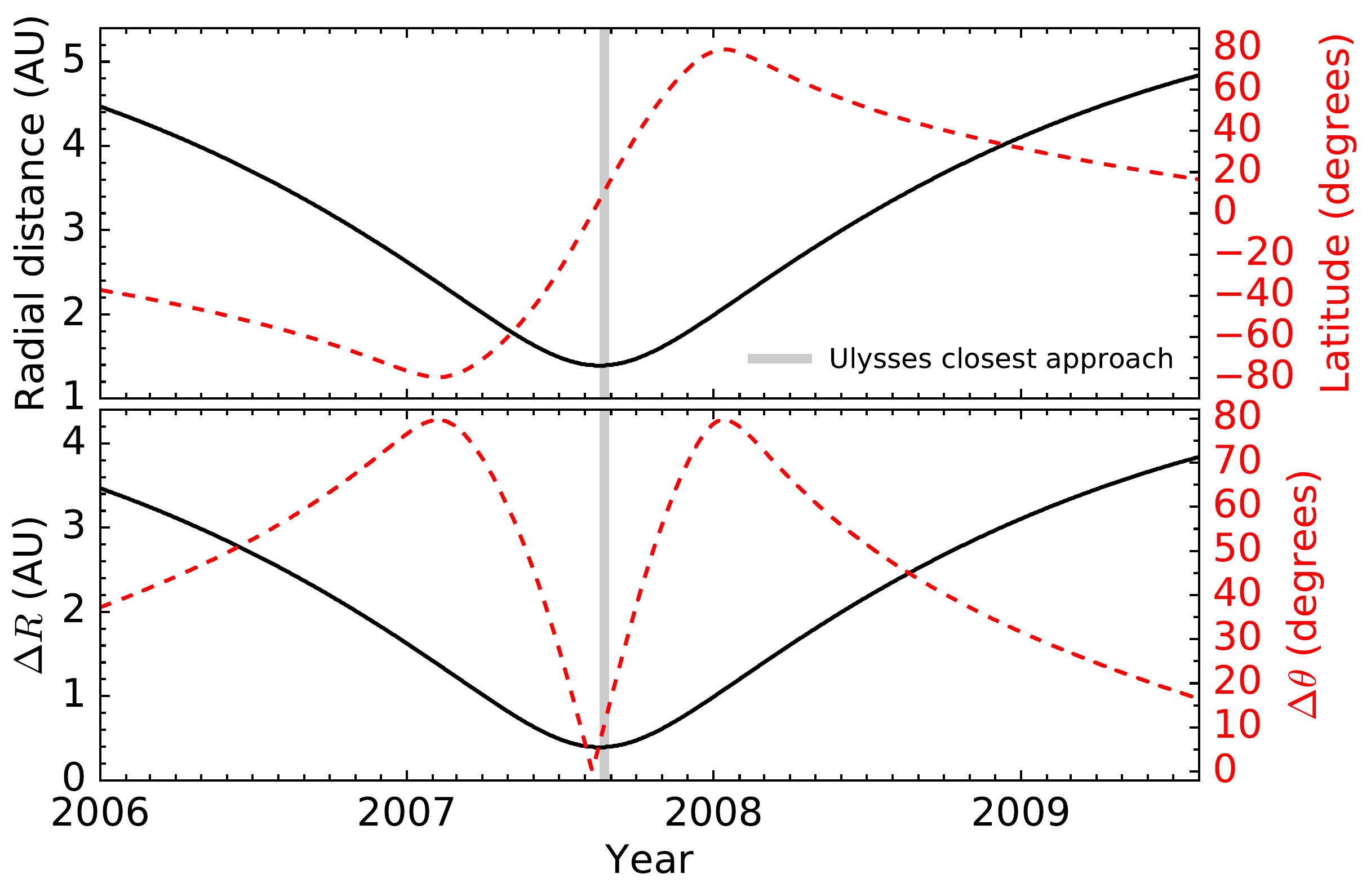}
  \end{center}
  \caption{{\it Top panel:}  Heliocentric distance [$r_U$] and latitude [$\theta_U$] of \emph{Ulysses} along its orbit are given by the solid black and dashed red lines, respectively, between 2006 and mid-2009.  {\it Bottom panel:}  The differences in distance [$\Delta r$] and latitude [$\Delta\theta$] between the positions of \emph{Ulysses} and PAMELA are given by the black and dashed red lines, respectively.  The vertical grey line indicates the time of \emph{Ulysses}' closest approach to Earth (see also \emph{e.g.} \citeauthor{Gieseleretal2013}, \citeyear{Gieseleretal2013}).\label{fig:prot_gradients_rthet}}
\end{figure}

Assuming that radial and latitudinal variations are separable and that these variations can be approximated by an exponential law, \mbox{Equation (\ref{eq:JuJeG})} can be rewritten as
\begin{equation}\label{eq:JuJeExp}
  J_U(P,t,r_U,\theta_U)\, =\, J_E(P,t,r_E,\theta_E)\; e^{G_r\Delta{r}}\; e^{G_\theta\Delta\theta},
\end{equation}
with $G_r$ and $G_\theta$ as before.  For an explanation on how corresponding rigidity channels from \emph{Ulysses}/KET and PAMELA are selected to ensure that data from similar rigidities are compared, see \citet{Desimoneetal2011}.

\subsection{Calculating Spatial Gradients}
To calculate $G_r$ and $G_\theta$ from observations, Equation (\ref{eq:JuJeExp}) is rewritten as
\begin{eqnarray}\label{eq:JuJeGrGt}
  \log\left(\frac{J_U}{J_E}\right) &=& G_r\Delta{r} + G_\theta\Delta\theta \nonumber\\
\frac{1}{\Delta{r}}\log\left(\frac{J_U}{J_E}\right) &=& G_r + G_\theta\frac{\Delta\theta}{\Delta{r}}.
\end{eqnarray}
Setting $X:={\Delta\theta}/{\Delta{r}}$ and $Y:={1}/{\Delta{r}}\log\left({J_U}/{J_E}\right)$, Equation (\ref{eq:JuJeGrGt}) is expressed as
\begin{equation}\label{eq:xygrgt}
  Y = G_r + G_\theta{X}.
\end{equation}
Assuming that $G_r$ and $G_\theta$ are independent of time and space, their respective values can be obtained from the offset and slope of a straight line fitted through the graph of {\it Y} against {\it X}.  Since this assumption is not very accurate for the time period of \emph{Ulysses}' fast latitude scan (FLS), between May and December of 2007, \citet{Gieseleretal2013} and \citet{Desimoneetal2011} studied the spatial gradients during the southern ascent, the FLS, and the northern descent phases of \emph{Ulysses}' trajectory separately and combined.  In our study, however, preceding yearly time periods were used, corresponding to the four year-end PAMELA spectra published by \citet{Adrianietal2013}.
\par
In determining $G_r$ and $G_\theta$, the intensity--time profiles of \emph{Ulysses}/KET and PAMELA are normalized to unity at the time of \emph{Ulysses}' closest approach to Earth in August 2007 (vertical grey line in Figure 5).  This is done as an attempt to minimize possible uncertainties in flux estimation related to the geometrical factor for \emph{Ulysses}/KET.
\par
\mbox{Figure 6} shows the normalized intensity measurements (left panels) from PAMELA (filled symbols) and \emph{Ulysses}/KET (open symbols) at rigidities between $460\,$MV and $1.90\,$GV, where the yearly intervals are colour-coded according to the legend (\citeauthor{Desimoneetal2011}, \citeyear{Desimoneetal2011};  \citeauthor{Adrianietal2013}, \citeyear{Adrianietal2013};  \citeauthor{Gieseleretal2013},  \citeyear{Gieseleretal2013};  Gieseler, 2014 (private communication)). The ongoing recovery of galactic CRs after August 2007 is clearly visible, even though the lowest number of sunspots were observed in 2007 and 2008 \citep[\emph{e.g.}][]{Heberetal2009}.  Both instruments show a similar temporal evolution in intensities, except that \emph{Ulysses}, at larger radial distances, measured higher relative intensities than PAMELA over most of its orbit.  In general, the differences between PAMELA and \emph{Ulysses}/KET can primarily be ascribed to spatial variations.
\begin{figure}
  \begin{center}
    \includegraphics[width=0.9\linewidth]{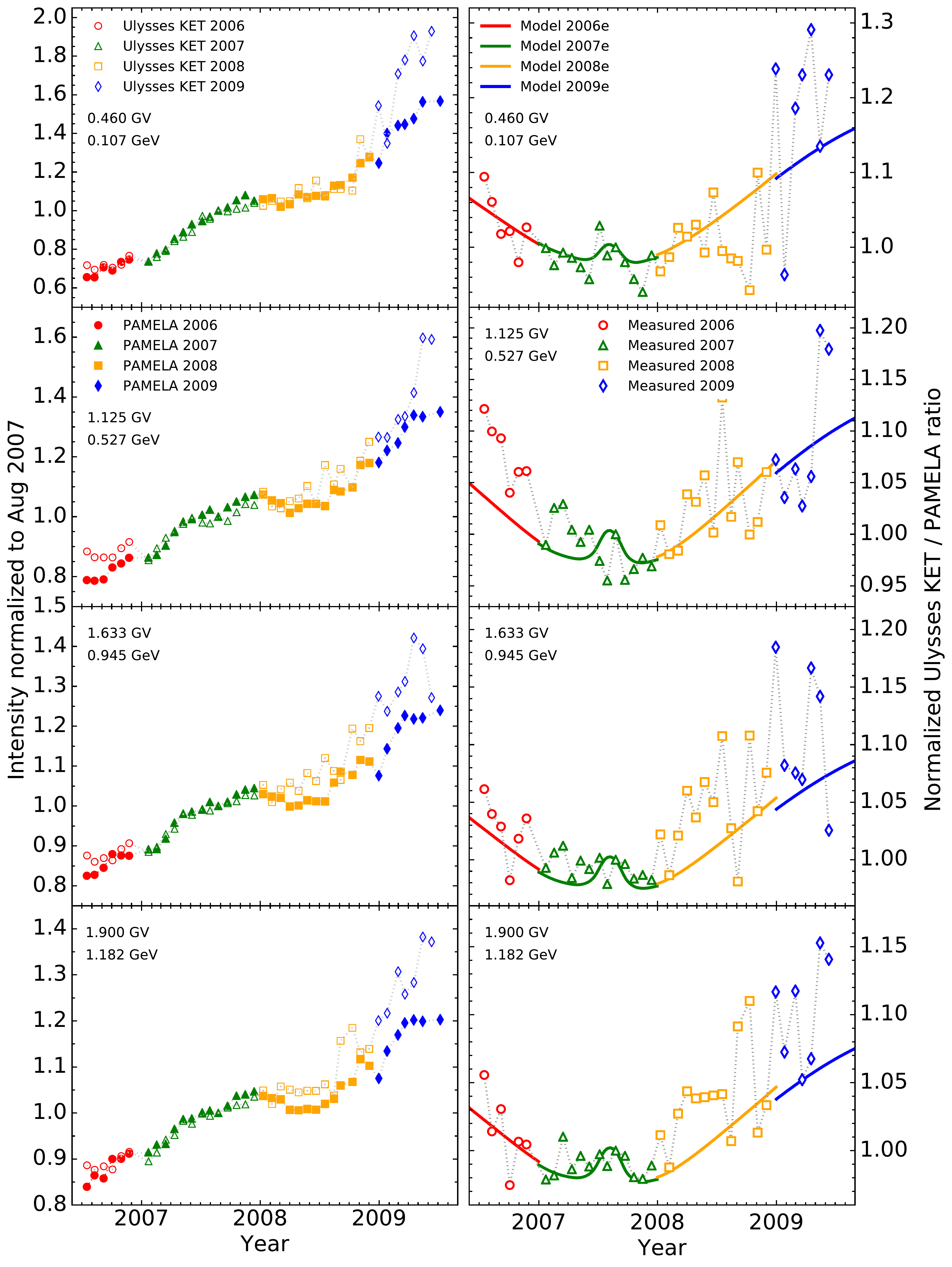}
  \end{center}
  \caption{{\it Left panels:}  The colour-coded time profiles of intensities measured by \emph{Ulysses}/KET (open symbols) and PAMELA (filled symbols), normalized at the time of \emph{Ulysses}' closest approach to Earth in August 2007.  Circles, triangles, squares, and diamonds represent measurements taken during 2006, 2007, 2008, and 2009, respectively.  {\it Right panels:}  Normalized \emph{Ulysses}/KET to PAMELA ratios [$J_u/J_E$], where observed and computed intensities are represented by the connected symbols and solid coloured lines, respectively.  From top to bottom, observed and computed intensities are shown for rigidities between $460\,$MV and $1.90\,$GV.\label{fig:prot_gradients_int_ratio}}
\end{figure}
\par
We show in the right panels of \mbox{Figure 6} the normalized observational (open symbols) and computed (lines) intensity ratios that appear as $J_U/J_E$ in Equation (\ref{eq:JuJeGrGt}).  From comparing the measured and computed ratios, it is clear that the model reproduces the overall trends in these ratios quite satisfactorily, with some differences between the model and measurements noticeable at $1.125\,$GV during 2006, and at $1.633\,$GV and $1.90\,$GV between 2008 and 2009.
\par
The effect of drifts is evident from the marked increase in intensity during 2007, when \emph{Ulysses} performed its FLS.  When positive particles drift inward mainly along the HCS during an $A < 0$ cycle, it results in higher intensities in the equatorial regions relative to the off-equatorial latitudes. This behaviour is confirmed by both observations and modelling as a characteristic feature of particle drifts; see also the review by \citet{Potgieter2013}.
\par
\begin{figure}[t]
  \includegraphics[width=1.0\linewidth]{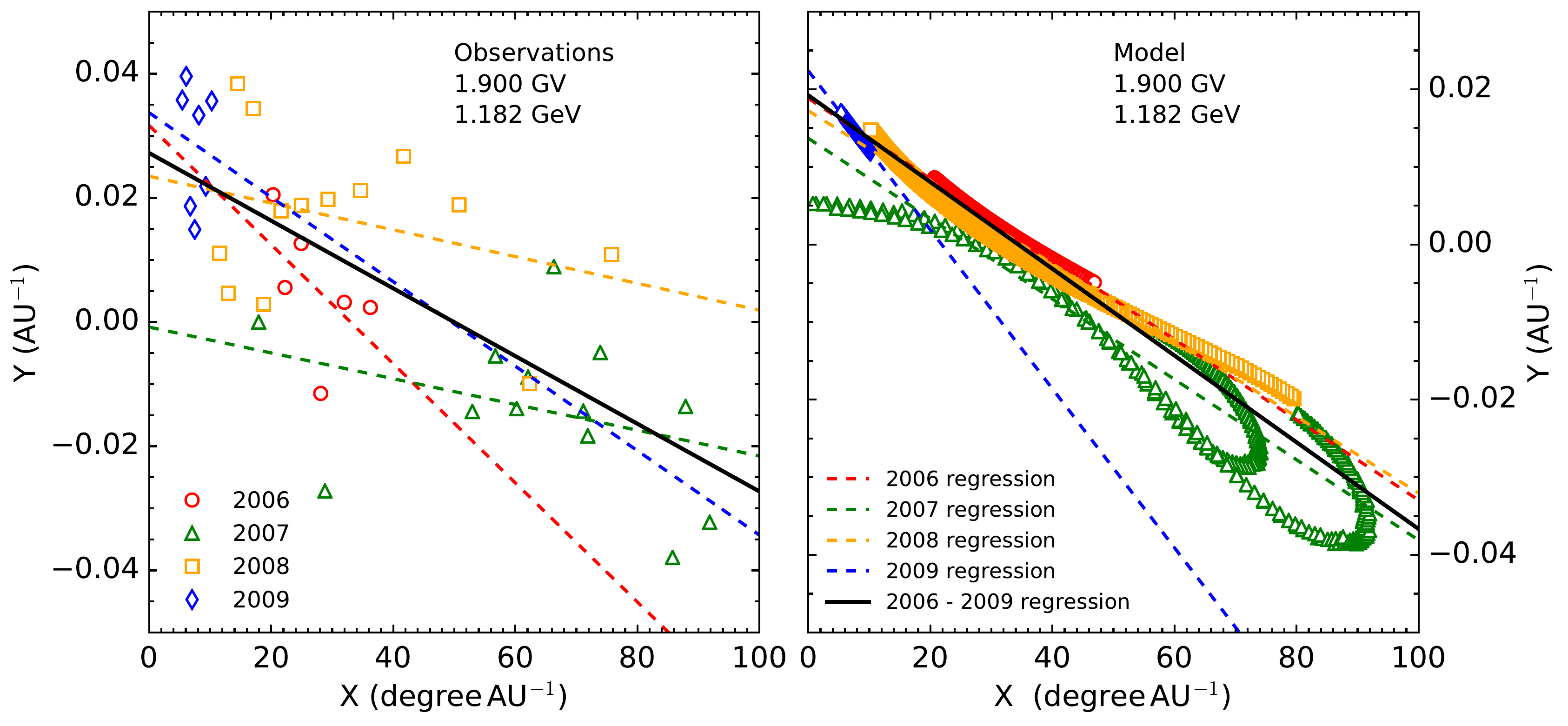}
  \caption{Quantities $Y$ and $X$ (from Equation (\ref{eq:xygrgt})) are plotted against each other for $1.90\,$GV protons.  The $X$- and $Y$-data and regression lines for the observational data and model data are given in the left and right panels respectively.  The coloured dashed lines are linear-regression fits that only take separate yearly data into account, while the solid black line represents the linear-regression fit for the combined dataset over the entire time period.  The values $G_r$ and $G_\theta$ are obtained from the offset and slope of the regression lines.\label{fig:prot_gradients_xy}}
\end{figure}
With $\Delta\theta$ and $\Delta{r}$ known, and $J_U/J_E$ calculated as in \mbox{Figure 6}, $Y$ is plotted against $X$, as shown in \mbox{Figure 7} for $1.90\,$GV proton measurements (left panel) and computed results (right panel).  A linear-regression fit (solid-black line) is calculated for a combination of all the yearly data points (symbols), so that $G_r$ and $G_\theta$ for the full time period can be obtained from the offset and slope of the fitted line, respectively.  Similarly, yearly values for $G_r$ and $G_\theta$ can be calculated from linear fits by considering only the corresponding yearly data points separately, as shown by the coloured dashed lines.  
The $X$ and $Y$ modelled values in the right panel of Figure 7 are better correlated than the observational values in the left panel. The resulting yearly gradients from the model are considered as an estimation range for temporal variation.
\begin{figure}[ht!]
  \begin{center}
    \includegraphics[width=0.85\linewidth]{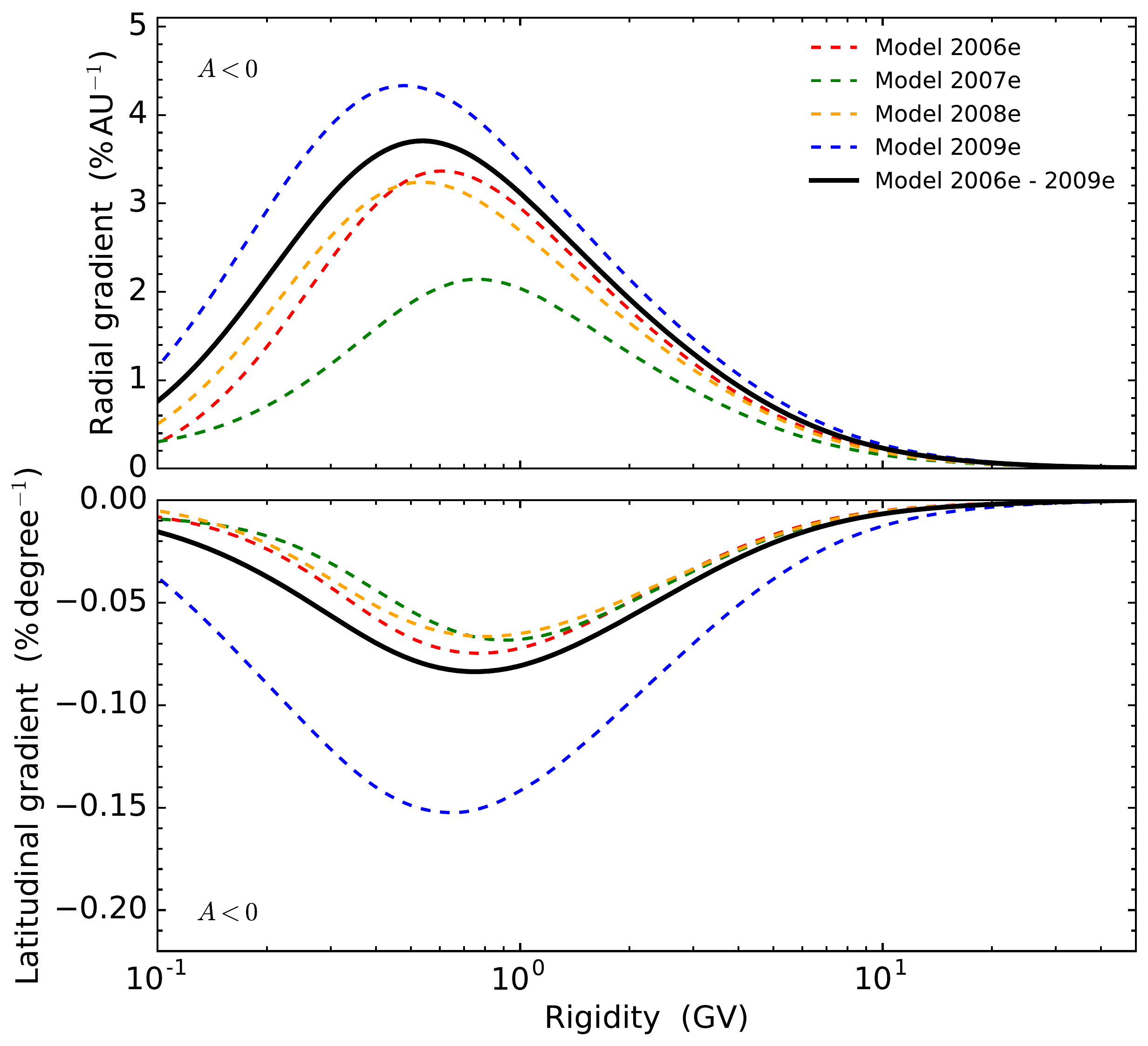}
  \end{center}
  \caption{{\it Top panel:}  Computed global radial gradients [$G_r$] between the positions of PAMELA (at the Earth) and along \emph{Ulysses}' orbit, from July 2006 to June 2009, which was during an $A < 0$ polarity cycle. {\it Bottom panel:} Computed global latitudinal gradients [$G_\theta$] similar to the top panel.  In both panels the year-end model gradients are given by the coloured dashed lines, while model gradients for the combined time period are given by the solid black line.} 
\end{figure}
\par
After the above procedure was applied to every rigidity step of the model solutions and to the available \emph{Ulysses}/KET and PAMELA observations, a comprehensive picture of radial and latitudinal gradients emerged.  
First, the modelling results are shown in Figure 8 as a function of rigidity, with $G_r$ and $G_\theta$ given in the top and bottom panels, respectively. 
This is provided for every year, and it serves to illustrate how these spatial gradients had developed in the inner heliosphere during this remarkable solar minimum modulation period.
The model evidently produces positive radial and negative latitudinal gradients, as expected from drift theory for protons in an $A < 0$ cycle. 
In the context of gradient predictions by drift-dominated models, $G_r$ for this study is larger than reported before, while $G_\theta$ is less negative, but still distinctively negative. 
According to this approach, the model predicts that the largest $G_r$ between the Earth and the position of \emph{Ulysses} occurred during 2009, with a maximum of $G_r=4.25\,$\%\,AU$^{-1}$ around $500\,$MV. 
The smallest $G_r$ is found for 2007 because $\Delta\theta$ varied significantly during 2007, when \emph{Ulysses} performed its FLS.  This makes our assumption that $G_r$ and $G_\theta$ are independent in terms of time and space less accurate.
\par
For the latitudinal gradients, the most negative value is found for 2009, with $G_\theta=-0.15\,$\%\,degree$^{-1}$ around $600\,$MV, while the least negative $G_\theta$ is found for 2007.  
A characteristic of the model is that these gradients decrease significantly below $\approx\,400\,$MV because drifts decrease toward lower rigidities, as shown in Figure 1. This illustrates that when drifts are reduced, the negative latitudinal gradients will dissipate. The reason for the difference in computed gradients for each of the four years is displayed in Figure 1 and was discussed in detail by \citet{PotgieterVosetal2014} and \citet{VosPotgieter2015}.
\par
\begin{figure}[ht!]
  \begin{center}
    \includegraphics[width=0.85\linewidth]{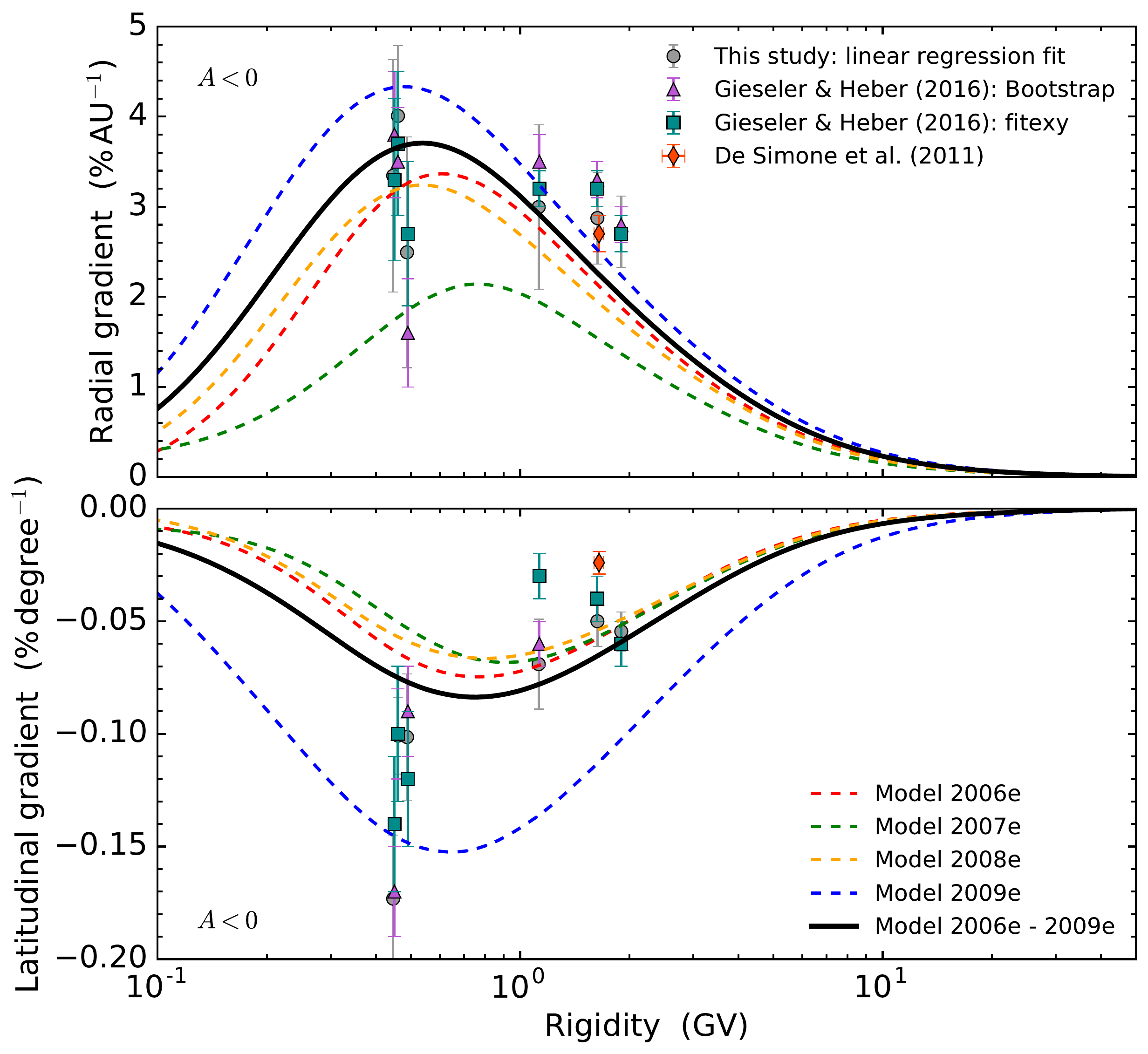}
  \end{center}
  \caption{{\it Top panel:} Global radial gradients from the model (in Figure 8) are compared to observational gradients calculated in this study (circles).  We also show two sets of calculated gradients from \citet{GieselerHeber2016}, who based their analysis on two statistical fitting algorithms, namely bootstrap (triangles) and the \textit{fitexy} function from Numerical Recipes (\citeauthor{Pressetal1996}, \citeyear{Pressetal1996};  squares).  The radial gradient calculated by \citet{Desimoneetal2011} is given by the diamond symbol.  {\it Bottom panel:} Similar to top panel, but for global latitudinal gradients.}
\end{figure}
In Figure 9 the modelling results from Figure 8 are compared to the observational gradients calculated in this study and to observational gradients from \citet{Desimoneetal2011} and the very recent study of \citet{GieselerHeber2016}. The latter investigation used a more sophisticated approach, producing two sets of results for global gradients based on two statistical fitting methods, as indicated in the figure. 
The original study of \citet{Desimoneetal2011} reported gradients for $1.6\,$GV -- $1.7\,$GV protons only (diamond symbols) and was based on preliminary proton spectra from PAMELA.  More recent observational gradients are based on the final proton spectra as published by \citet{Adrianietal2013}. 
\par
Our calculated observational gradients are consistent with $G_r$ and $G_\theta$ values reported by other authors, taking error bars into account. Most important, from our modelling point of view, good agreement is found between the observational global gradients (symbols) and modelled global gradients (lines) over the rigidity range considered here. The global modulation produced by the model is compatible with the global modulation observed during the period from 2006 to 2009. Given that a complicated analysis of observations is required for this type of study, as discussed in detail by \citet{GieselerHeber2016}, we refrain from speculating why some differences are present. 

\section{Discussion and Conclusions}
The reported PAMELA proton spectra observed between mid-2006 and the end of 2009 enabled us to perform a comprehensive modelling study of CR modulation during this unusual solar minimum period, as reported by \citeauthor{PotgieterVosetal2014} (\citeyear{PotgieterVosetal2014}, \citeyear{PotgieterVosetal2015}), and extended by \citet{VosPotgieter2015}. 
A newly constructed vLIS was used as an input spectrum that takes into account recent \emph{Voyager}-1 observations at low energies. Based on these studies, the conclusion was made that the minimum-modulation period of 2009 was relatively more diffusion-dominated instead of being drift-dominated as previous $A < 0$ polarity cycles seem to have been. See also the assessment of solar minimum spectra by \citet{StraussPotgieter2014}, and their conclusion that the 2009 PAMELA proton spectrum was the highest recorded in the space era. The studies mentioned here illustrate that drifts played a notable role, especially because the HMF had decreased significantly until the end of 2009, in contrast to the moderate decreases observed during previous minimum periods.
\par
For this article, we began by using the model of \citet{VosPotgieter2015}, with the parameters tuned to reproduce four year-end  PAMELA proton spectra.  We applied the model to compute the radial dependence of the proton spectrum throughout the heliosphere for 2009. Corresponding radial profiles were also computed for each year, from 2006 onwards, along the \emph{Voyager}-1 trajectory, and compared to available observations. 
\par
It is found that the computed intensity levels are in good agreement with solar minimum observations from \emph{Voyager}-1 at multiple energies. In addition, the model also reproduces the steep intensity increases observed when \emph{Voyager}-1 crossed the HP region.  This increase was comprehensively simulated with an independent model by \cite{Luoetal2015}. In this context, we conclude that our model gives a most reasonable presentation of the CR radial profiles, from the Earth to the HP, for 2006 to 2009.
\par
Simultaneous observations from \emph{Ulysses}/KET and PAMELA, between July 2006 and June 2009, were made available, allowing for a study of the radial and latitudinal gradients in the inner heliosphere. To ensure a meaningful comparison between our modelling results and observations, we applied an empirical method, similar to that of \citet{Desimoneetal2011} and \citet{GieselerHeber2016}, to our modelling results at the Earth and along \emph{Ulysses}' orbit for this period.  We find good agreement between the computed values from the model and those calculated from observations, for both the radial and latitudinal gradients. 
\par
We conclude that the model also gives a most reasonable representation of spatial gradients in the inner heliosphere for 2006 to 2009. Our computations reflect that drifts influenced CR modulation during this unusual solar minimum, so that the notion that drifts were unimportant during the recent solar minimum (\emph{e.g.} \citeauthor{Cliveretal2013}, \citeyear{Cliveretal2013}) is not supported. Even though the observable effects of drifts are somewhat suppressed by the excess diffusion, drifts still maintained a strong presence, as explained by \citet{PotgieterVosetal2014}.
We emphasize that the drift effects shown here with a model tuned to the special conditions during the 2006 to 2009 solar minimum are indeed weaker then previous predictions of drift-dominated models for $A < 0$ cycles; 
see also the review by \citet{Potgieter2014b}. Evidently, nobody had foreseen that the minimum-modulation conditions for the $A < 0$ cycle of 2009 would be so different and unusual.
\par
It follows from our simulations for the period 2006 to 2009 that drifts had the strongest effect on global latitudinal gradients in the inner heliosphere (as covered by \emph{Ulysses}) in the rigidity range of $0.6\,$GV -- $1.0\,$GV, but this subsided rather quickly with decreasing rigidity. The most negative latitudinal gradient is computed for 2009, with a value of $G_\theta=-0.15\,$\%\,degree$^{-1}$ around $600\,$MV. The highest value for the global radial gradient in the inner heliosphere is $G_r=4.25\,$\%\,AU$^{-1}$ around $500\,$MV in 2009 as well.

\section*{Acknowledgments}
The authors express their gratitude to the South African National Research Foundation (NRF) for providing partial funding. EE Vos also thanks the South African National Space Agency (SANSA) for partial financial support during his PhD studies. 
They thank Jan Gieseler for insightful discussions during his visit to South Africa and for providing them with updated \emph{Ulysses} observations. They also thank Mirko Boezio, Valeria di Felice, and Riccardo Munini as members of the PAMELA team for many insightful discussions on the PAMELA data.



\end{document}